# Efficient Quantum Transduction Using Anti-Ferromagnetic Topological Insulators

Haowei Xu[1], Changhao Li[1,2], Guoqing Wang[1,2], Hao Tang[3],
Paola Cappellaro[1,2,4, †], and Ju Li[1,3, ‡]

[1] Department of Nuclear Science and Engineering, Massachusetts Institute of Technology, Cambridge, Massachusetts 02139, USA

[2] Research Laboratory of Electronics, Massachusetts Institute of Technology, Cambridge, MA 02139, USA

[3] Department of Materials Science and Engineering, Massachusetts Institute of Technology, Cambridge, Massachusetts 02139, USA

[4] Department of Physics, Massachusetts Institute of Technology, Cambridge, MA 02139, USA

\* Corresponding authors: † pcappell@mit.edu,   ‡ liju@mit.edu

## Abstract

Transduction of quantum information between distinct quantum systems is an essential step in various applications, including quantum communications and quantum computing. However, mediating photons of vastly different frequencies and designing high-performance transducers are highly nontrivial, due to multifaceted and sometimes conflicting requirements. In this work, we first discuss some general principles for quantum transducer design, and then propose solid-state anti-ferromagnetic topological insulators to serve as particularly effective transducers. First, the anti-ferromagnetic order can minimize detrimental influences on nearby quantum systems caused by magnetic interactions. Second, topological insulators exhibit band-inversion, which can greatly enhance their optical responses. This property, coupled with robust spin-orbit coupling and high spin density, results in strong nonlinear interaction in magnetic topological insulators, thereby substantially improving transduction efficiency. Using $MnBi_2Te_4$ as an example, we discuss the potential experimental realization of quantum transduction based on magnetic topological materials. Particularly, we showcase that quantum transduction efficiency exceeding 90% can be achieved with modest experimental requirements, while the transduction bandwidth can reach the GHz range. The strong nonlinear photonic interactions in magnetic topological insulators can find diverse applications besides quantum transduction, such as quantum squeezing.

**Introduction**. Quantum information processing systems typically operate in distinct frequency domains. For example, superconducting qubits, one of the leading platforms in quantum computation, work at microwave (MW) frequencies. Meanwhile, photons with infrared (IR) to visible frequencies can serve as flying qubits for long-distance communication. Therefore, efficient quantum transduction between different frequency domains is crucial to harness the advantages of different platforms in hybridized quantum systems and quantum networks.

We specifically focus on the transduction between MW and IR photons. An ideal quantum transducer should exhibit strong interaction with both MW and IR photons, so that high transduction efficiency and wide bandwidth can be enabled. Meanwhile, it is also necessary to minimize any adverse influence on nearby quantum systems resulting from the coupling with the transducer. Diverse transduction schemes have been explored, involving neutral atoms [1,2], rare-earth-doped crystals [3,4], optomechanical devices [5–7], electro-optics [6–8], and magnons [9,10], etc. Unfortunately, it is a nontrivial task to identify a high-performance transducer. For example, while Rydberg atoms have relatively strong interaction with MW photons, their large polarizabilities can also lead to decoherence and even destroy the superconducting states of nearby superconducting qubits [11,12]. Electro-optical devices, on the other hand, suffer from weak nonlinearities typically below $10^3$ pm/V [13].

Magnetic topological insulators (MTIs) [14–17] are a class of magnetic materials with nontrivial electronic band topology, which have been attracting wide interest in condensed matter physics. In MTIs, the combination of the nontrivial electronic band topology and magnetism leads to a wealth of novel properties, such as the quantum anomalous Hall effect [16] and axion electrodynamics [17]. These unique properties suggest novel applications of MTIs. In this work, we propose that MTIs, particularly anti-ferromagnetic topological insulators (AFMTIs), hold great potential in quantum information science as well, thanks to their enhanced linear and nonlinear optical response. Specifically, we will demonstrate that AFMTI can serve as efficient quantum transducers, which could enable high transduction efficiency and wide bandwidth exceeding GHz even at the single-photon level.

Magnetic materials have ultra-large number density ($\sim 10^{28}$ m$^{-3}$) of magnetic moments, and their inherent collective spin excitation (magnon) frequencies typically fall in the MW range [18–20]. These properties suggest that magnetic materials can have strong coupling to MW photons [9,10].

Additionally, if anti-ferromagnetic (AFM) materials are used as the transducer, the influence on nearby quantum systems can be minimized, because AFM materials have vanishing total magnetic moments. Indeed, these features have elicited great interest in AFM spintronics [18,19].

Moreover, the nontrivial electronic band topology of MTIs can give rise to an additional boost in quantum transduction performance. Topological insulators feature band inversion [21,22], whereby the energy ordering of normal valence and conduction bands is inverted in certain regions of the Brillouin zone. The band inversion usually leads to stronger hybridization between valence and conduction band wavefunctions, which in turn results in larger Berry curvature and stronger coupling with visible and IR photons [23–26]. Meanwhile, topological insulators typically possess heavy elements and thus strong spin-orbit coupling (SOC), which accelerates the conversion between spin and orbital dynamics and further improves the transduction efficiency in MTIs. Synergistically, these features contribute to the proposed performance of MTI quantum transducers.

In the following, we will first discuss some general principles to improve transduction efficiency. Then, we will introduce the mechanism of quantum transduction in magnetic materials, and explain why AFMTIs can be efficient transducers. Using $MnBi_2Te_4$ [27–29] as an example, we will demonstrate the enhanced coupling strength with MW and IR photons in AFMTIs. Specifically, the intrinsic second-order $\chi^{(2)}$ nonlinearity of $MnBi_2Te_4$ can reach $10^6$ pm/V, orders of magnitude larger than that of typical electro-optical materials [13] and topologically trivial magnetic materials [9,10]. Thereafter, we will discuss several issues pertinent to the potential experimental realization of quantum transductions based on AFMTIs. We will show that quantum transduction using $MnBi_2Te_4$ as the transducer can achieve a transduction efficiency of over 90% with modest experimental requirements. Besides, the bandwidth of the transduction can potentially reach GHz, nearly two orders of magnitude larger than the typical bandwidth based on other transduction schemes [30]. These results highlight the significance of large nonlinearities of AFMTIs for overcoming the hurdles in achieving efficient quantum transduction.

**Design principles for efficient transducers.** The purpose of the transduction process is to convert MW photons to IR photons (or vice versa), whose frequencies are denoted as $\omega_{MW}$ and $\omega_{IR}$, respectively. An IR pumping laser with a frequency $\omega_{pump} = \omega_{IR} - \omega_{MW}$ is applied, so that energy conservation can be satisfied. Note that the transduction scheme described in this work is based on three-wave mixing. It is also possible to use four- or more-wave mixing, whereby additional MW or IR pumps should be applied, or opto-mechanical systems, whose performance has been significantly improved recently [6,31]. To achieve efficient quantum transduction, the transducer should have strong coupling with both MW and IR photons, whose frequencies are drastically different by 4-5 orders of magnitude. Establishing strong coupling between quantum degrees of freedom (DoF), such as photons and spins, generally requires resonance (frequency matching) conditions. For example, to trigger efficient spin flip-flop transitions, the MW photon needs to be resonant with the spin transition energy. However, a single DoF cannot be resonant with two vastly different frequencies simultaneously. Actually, this leads to an intrinsic limit on the coupling strength of the electro-optical effect from pure electronic responses [32], whereby the MW and the IR/visible photons are both coupled to the electronic orbital transitions ($\sim 1$ eV) in semiconductors. The MW photons ($\sim 0.1$ meV) are far off-resonance, leading to weak responses. Indeed, in materials with soft phonon modes, such as $BaTiO_3$ [33,34], the major contribution to the electro-optical responses comes from ionic (i.e., phonon) responses [13], thanks to the relatively smaller detuning between phonons and MW photons.

A potentially better strategy is to employ a transducer with two DoFs designed to be (nearly) resonant with MW and IR photons, respectively. These two DoFs are then coupled by a specific internal interaction within the transducer. For example, using Rydberg atoms, the MW photons are coupled with the transitions between two highly excited Rydberg states. By judiciously selecting the principal quantum numbers of the Rydberg states, nearly resonant conditions can be satisfied, which can significantly improve the coupling strength. Unfortunately, this also leads to some detrimental effects, such as large polarizabilities [30].

Similarly, the transduction scheme based on rare-earth-ion doped crystals [3,4] or magnons [9,10] utilizes the two intrinsic DoFs of electrons, namely spin and orbital dynamics. These two DoFs are internally coupled by SOC, which is ubiquitous in atoms, molecules, and solid-state systems. The orbital excitation energy of electrons is usually on the order of 1 eV, facilitating the coupling with photons in the visible or IR range. To utilize the spin DoF for the coupling with MW photons, it is necessary to have unpaired electron spins (magnetic moments), which can result from dopants, such as rare-earth-ions, in crystals. In terms of coupling strength, a more favorable option is to use pristine magnetic materials [9,10], where the number density of electron magnetic moments can reach $10^{28}$ m$^{-3}$. Besides, external magnetic fields are not necessary for sustaining the spin splitting in magnetic materials, which can be advantageous as well. In ferromagnetic materials, the spin excitation (magnon) frequency $\omega_m$ is typically on the order of GHz and is close to MW frequencies, but the macroscopic magnetic moments and the resultant magnetic fields can adversely affect the performance of the quantum objects (e.g., superconducting qubits) adjacent to the transducer [9,10].

In this regard, we suggest using AFM materials, whose net macroscopic magnetic moment is zero. However, a potential concern arises from the high magnon frequencies $\omega_m$ in AFM materials (up to the THz range [18,19]), which can lead to relatively large detuning from the frequencies of MW photons. Therefore, soft AFM materials with relatively low magnon frequencies (down to 100 GHz) would be more favorable. Furthermore, it is crucial to establish a general approach for selecting or designing AFM materials that can strongly interact with IR and/or MW photons, so that the overall transduction efficiency can be improved.

The discussions above lead us to the proposal of using AFMTIs as the transducer. As we will explain in more detail later, compared with topologically trivial AFM materials, AFMTIs have enhanced coupling strength with IR photons, thanks to their band inversion. AFMTIs also possess strong intrinsic SOC, resulting in rapid internal conversion between spin and orbital dynamics, which is beneficial for quantum transduction. Consequently, the transduction efficiency can be strong despite the relatively large frequency detuning between AFM magnons and MW photons. In the following, we will argue that AFMITs can serve as efficient quantum transducers, and their transduction performance can be orders of magnitude better than topologically trivial AFM materials, such as $Cr_2O_3$.

**Transduction mechanism in anti-ferromagnetic materials.** We first examine the mechanism of transduction using AFM materials. Although magnons in AFM materials are collective spin excitations, we will treat every unit cell individually for the sake of simplicity. This approach is valid since photons only interact with magnons with wavevectors $\boldsymbol{k} \approx 0$. SOC will only be considered for the interconversion between spin and orbital dynamics and will be disregarded elsewhere. The quantum state of electrons can be labeled with $|\mathcal{N}, \alpha\rangle$, where the electron wavevector $\boldsymbol{k}$ is omitted, $\mathcal{N} = 0,1,2,...$ denotes the number of magnons, while $\alpha$ denotes the orbital state of the electrons. The energy of $|\mathcal{N}, \alpha\rangle$ is labeled as $E_{|\mathcal{N},\alpha\rangle}$. We assume the electron is initially on the $|0, v\rangle$ state with $\alpha = v$ a valence orbital state. The electron, as the transducer, shall return to this initial state after the whole transduction process is finished. Note that in magnetic materials, the localized magnetic moment mostly comes from the electron spins, instead of orbital angular momenta [35]. We will thus use spin and magnetic moment interchangeably hereafter.

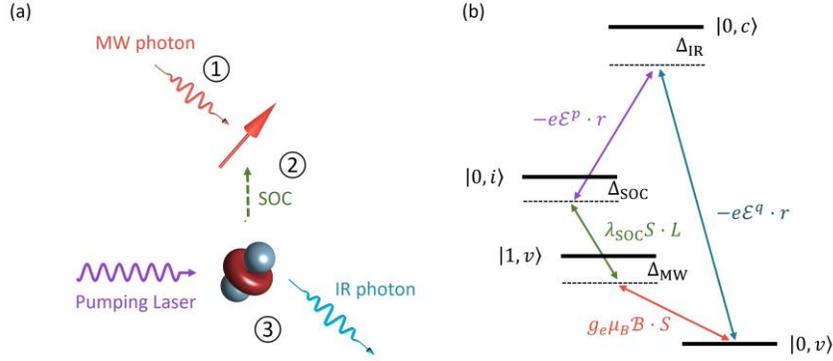

**Figure 1**. Schematic diagram showing the mechanism of the transduction based on MTIs (the MW→IR process). MW and IR photons interact with the spin and orbital degrees of freedom of the electrons, respectively, which are internally coupled by SOC. A pumping laser is applied so that energy conservation can be fulfilled. (b) Energy level diagram of the transduction process. See the main text for a description of the scheme and labels.

To facilitate further discussions, we separate the transduction process into three steps (Figure 1a). First, the magnetic field $\mathcal{B}$ of the MW photon interacts with the electron spin $S$ through the Zeeman interaction $H_Z = g_e \mu_B S \cdot \mathcal{B}$, where $g_e$ is the electron $g$-factor and $\mu_B$ is the Bohr magneton. Due to the Zeeman interaction, one magnon can be excited by an MW photon and the electron thus jumps to the $|1, v\rangle$ state. The strength of this first step can be obtained from perturbation theory, and can be characterized by $\xi_{\mathrm{MW}} = \frac{\langle 0,v|H_Z|1,v\rangle}{\Delta_m}$, where $\Delta_m \equiv E_{|1,v\rangle} - E_{|0,v\rangle} - \omega_{\mathrm{MW}} = \omega_m - \omega_{\mathrm{MW}}$ is the frequency mismatch between the MW photon and the magnon.

Next, SOC induces a $|1, v\rangle \to |0, i\rangle$ transition, whereby the electron spin DoF returns to the ground state ($\mathcal{N} = 0$), while the orbital DoF transits from $|v\rangle$ to an intermediate state $|i\rangle$. The strength of this second step is characterized by $\xi_{\mathrm{SOC}} = \frac{\langle 1,v|H_{\mathrm{SOC}}|0,i\rangle}{\Delta_{\mathrm{SOC}}}$, where $H_{\mathrm{SOC}} = \lambda_{\mathrm{SOC}} S \cdot L$ is the SOC Hamiltonian, $\lambda_{\mathrm{SOC}}$ is the SOC strength, $L$ is the angular momentum operator, while $\Delta_{\mathrm{SOC}} \equiv E_{|0,i\rangle} - E_{|1,v\rangle} - \omega_{\mathrm{MW}}$ is the energy detuning. Here the SOC is considered a static interaction. Generally, the numerator $\langle 1, v|H_{\text{SOC}}|0, i\rangle$ has a similar order of magnitude for all possible $i$. Hence, $\xi_{\text{SOC}}$ can be maximized when $\Delta_{\text{SOC}}$ is minimized. This condition is satisfied when $i = v$, whereby one has $\Delta_{\text{SOC}} \sim 100$ GHz $\sim 1$ meV, corresponding to MW frequencies, otherwise one usually has $\Delta_{\text{SOC}} \sim 1$ eV for $i \neq v$. In the case of $i = v$, the MW photon first excites a magnon, that is, the electron spins are rotated from their ground state orientations by the MW photons (step 1 in Figure 1). Then, SOC rotates the electron spins back to their ground state orientations (step 2 in Figure 1). During this process, the electron remains in the same orbital state $v$.

For the third step of the transduction process, two IR fields $p$ and $q$ interact with the electron orbital motion via the electric dipole interaction $H_{\text{d}}^{p\,(q)} = -e\mathcal{E}^{p\,(q)} \cdot r$, where $r$ is the electron position operator, and $\mathcal{E}^{p\,(q)}$ is the IR electric field. For definiteness, the $p$ and $q$ fields will hereafter correspond to the pumping laser and the transduced IR photon, respectively (Figure 1). Specifically, the $|0, i\rangle$ electron jumps to another intermediate state $|0, c\rangle$ via $H_{\text{d}}^p$, whereby a $p$-photon is absorbed, and then returns to the initial state $|0, v\rangle$ via $H_{\text{d}}^q$, whereby a $q$-photon is emitted. The strength of this two-photon process can be characterized by $\xi_{\text{IR}} = \sum_c \frac{\langle 0,i|H_{\text{d}}^p|0,c\rangle\langle 0,c|H_{\text{d}}^q|0,v\rangle}{\Delta_{\text{IR}}}$ with $\Delta_{\text{IR}} \equiv E_{|0,c\rangle} - E_{|0,v\rangle} - \omega_{\text{IR}}$, where $\omega_{\text{IR}}$ is the frequency of the $q$-photon. To the leading order of the analysis, we can also take $\omega_{\text{IR}}$ to be the frequency of the $p$-photon due to the small frequency difference between the two IR fields.

Altogether, the MW to IR transduction process is: (1) incident MW photons (virtually) excite magnons, i.e., spin excitations; (2) the spin excitations are converted to orbital excitations through SOC; (3) following the laser-pumped transitions, the orbital excitations decay, resulting the emission of IR photons (Figure 1). In principle, these three steps have equal status and can occur simultaneously. The separation into three steps here should be regarded as artificial, which leads to the fact that $\xi_{\text{MW}}$ and $\xi_{\text{SOC}}$ are unitless, while $\xi_{\text{IR}}$ has a unit of energy. A more rigorous treatment can be formulated using third-order perturbation theory [36]. The overall coupling strength of the transduction can be expressed as (see Section 1 of Ref. [37] for details, which also contains Refs. [5,38,47–51,39–46])

$$\mathcal{G} \approx \frac{N_s g_e \mu_B S\, \mathcal{B} \cdot \lambda_{\text{SOC}} \cdot \varepsilon_0 \chi_{\text{r}}(\omega_{\text{IR}}) V_{\text{cell}} \mathcal{E}^p \mathcal{E}^q}{\Delta_m \Delta_{\text{SOC}}} \tag{1}$$

Here $\varepsilon_0$ and $\chi_r(\omega_{IR})$ are the vacuum permittivity and the optical susceptibility of the host material, respectively. Meanwhile, $N_s$ and $V_{cell}$ are the total number of magnetic moments and the volume of the unit cell. In AFM materials, the magnetic moments are fully concentrated, and one has $N_s V_{cell} \sim V_0$ with $V_0$ the volume of the crystal. Note that $N_s$ is the number of spins that are simultaneously activated by the MW field $\mathcal{B}$ and the two IR fields $\mathcal{E}^p$ and $\mathcal{E}^q$. While Eq. (1) omits the influence of the spatial profiles of these three fields and assumes perfect mode overlap between them, we will later introduce a filling factor that accounts for the imperfect mode overlap.

From Eq. (1), one can observe that the coupling strength $\mathcal{G}$ is determined by the product of the strength of the three steps shown in Figure 1, which are represented by $N_s g_e \mu_B S\mathcal{B}$, $\lambda_{SOC}$, and $\varepsilon_0 \chi_r(\omega_{IR}) V_{cell} \mathcal{E}^p \mathcal{E}^q$, respectively. In addition, $\mathcal{G}$ is also influenced by $\Delta_m$ and $\Delta_{SOC}$. For AFM magnons, one typically has $\Delta_m, \Delta_{SOC} \gtrsim 100 \text{ GHz}$, which is well above the magnon linewidth [18,19]. $\Delta_m$ and $\Delta_{SOC}$ can potentially be reduced by applying externally static magnetic fields $\mathcal{B}_0$, which, however, may lead to unwanted influence on nearby quantum systems. Fortunately, $\mathcal{G}$ is sufficiently strong even if $\mathcal{B}_0 = 0$, as demonstrated later.

Additionally, we would like to remark that the transduction scheme described above can be used in centrosymmetric materials. Specifically, under spatial inversion operation, $\mathcal{G} \sim \mathcal{B}\mathcal{E}^2$ in Eq. (1) is invariant, as $\mathcal{B}$ and $\mathcal{E}$ get a $+1$ and $-1$ sign, respectively. In comparison, the electro-optical effect requires inversion symmetry breaking, because its coupling strength $\mathcal{G}_{EO} \sim \mathcal{E}^3$ undergoes a sign change under spatial inversion, hence it must be zero in centrosymmetric materials.

**Enhanced coupling strength in material topological insulators.** To improve the transduction performance, it is essential to enhance $\mathcal{G}$. This involves optimizing each of the three steps described above. Typically, the electron $g$-factor is close to 2 in magnetic materials [35,40,52–54]. Hence, to improve the strength of the first step (Zeeman interaction), one must increase the number of electron spins $N_s$ that participate in the interaction. This highlights the advantage of using magnetic materials, where the magnetic moments are fully concentrated [10], resulting in large $N_s$.

In contrast, the strengths of the second and third steps, which are determined by the SOC strength $\lambda_{SOC}$ and the optical susceptibility $\chi_r(\omega_{IR})$, can vary significantly in different materials. Notably, $\lambda_{SOC}$ and $\chi_r(\omega_{IR})$ in many scenarios are independent of each other. A strong SOC does not necessarily indicate strong optical responses, and vice versa [55]. Actually, some materials possessing strong optical responses, such as two-dimensional graphene [56] or MoS$_2$ [57], have weak SOC. Is there a systematic strategy to improve $\lambda_{\text{SOC}}$ and $\chi_r(\omega_{\text{IR}})$ simultaneously? MTIs [14,15] provide a good solution. Strong SOC is usually a necessary condition for non-trivial electronic band topology, and thus topological insulators possess heavy elements and strong SOC [21,22]. Meanwhile, the electronic band inversion in topological insulators leads to significant wavefunction hybridization between the valence and conduction bands. This results in faster interband transition rates and thus stronger bulk optical responses [23–25], which has been verified experimentally [26].

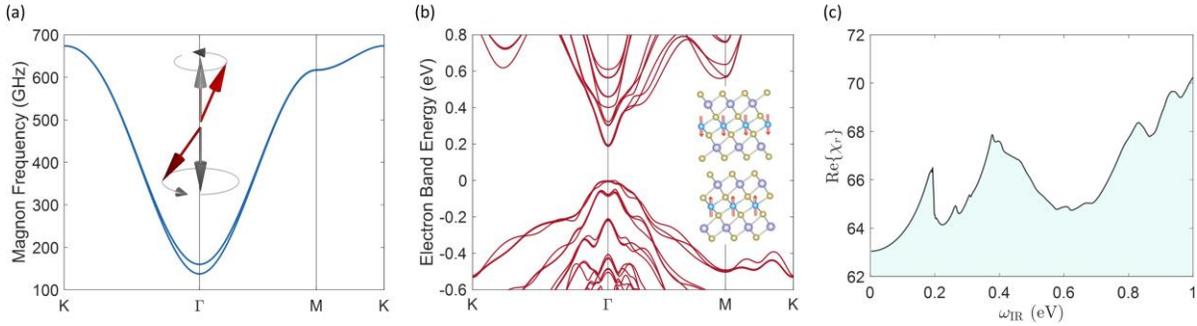

**Figure 2**. (a) Magnon dispersion of MnBi$_2$Te$_4$. The inset shows the spin precession in AFM materials when magnons are excited. Spins (red arrows) precess around the ground state (grey arrows, no magnons). (b) Electronic band structure of MnBi$_2$Te$_4$. The inset shows the atomic structure of MnBi$_2$Te$_4$. Blue: Mn; Purple: Bi; Brown: Te. The arrows on Mn atoms indicate the magnetic moments with AFM configuration. (c) The real part of the optical susceptibility of MnBi$_2$Te$_4$.

Next, we will use a well-known AFMTI, namely MnBi$_2$Te$_4$ [27–29], as an example to demonstrate the potential of AFMTIs as quantum transducers. MnBi$_2$Te$_4$ has a layered structure (inset of Figure 2b), and on each layer, magnetic Mn atoms are sandwiched by Bi and Te atoms in the sequence of Te–Bi–Te–Mn–Te–Bi–Te. A minimal Hamiltonian for the local magnetic momentum ($S = 5/2$) of Mn atoms is [39,40]

$$\mathcal{H}_s = J_1 \sum_{\langle ij \rangle} \boldsymbol{S}_i \cdot \boldsymbol{S}_j + J_c \sum_{\langle ij \rangle'} \boldsymbol{S}_i \cdot \boldsymbol{S}_j + D_z \sum_i (S_i^z)^2 + \gamma_e \sum_i \boldsymbol{B_0} \cdot \boldsymbol{S}_i. \quad (2)$$

Here $i,j$ label the Mn atoms, while $\langle ij \rangle$ and $\langle ij \rangle'$ indicate intra- and inter-layer nearest neighbors with spin-spin coupling strength $J_1 = -0.23 \text{ meV}$ and $J_c = 0.065 \text{ meV}$, respectively [39,40]. $D_z = -0.15 \text{ meV}$ characterizes the easy-axis magnetic anisotropy along $z$ axis. With external magnetic field $\mathcal{B}_0 = 0$, the intralayer exchange coupling between Mn atoms is ferromagnetic, while different layers are stacked with an AFM order. The magnon band structure obtained by linearizing Eq. (2) is shown in Figure 2a. The magnon frequency at the Γ-point is $\omega_m \approx 150 \text{ GHz}$, which is relatively low among AFM materials. This is advantageous compared with other hard AFM materials, since the detuning $\Delta_m$ from the MW photons would be smaller for $\omega_{MW} \sim 10 \text{ GHz}$. Also, magnon relaxation due to e.g., magnon-magnon scatterings and magnon-photon scatterings can potentially be weaker [58,59] in soft magnetic materials such as MnBi$_2$Te$_4$. This results from (1) smaller magnon-magnon coupling strength and (2) smaller phase space of magnons/phonons in which the energy conservation law can be satisfied during the scattering process. Note that MnBi$_2$Te$_4$ has relatively high symmetry, leading to degenerate magnon modes when the external magnetic field is $\mathcal{B}_0 = 0$. In this case, one can use circularly polarized microwaves to selectively excite certain magnon modes.

While Mn atoms provide magnetic moments, the heavy elements Bi and Te contribute to strong SOC in MnBi$_2$Te$_4$. According to our *ab initio* calculations, the SOC strength relevant to the transduction process is $\lambda_{SOC} \sim 47 \text{ meV}$ in MnBi$_2$Te$_4$ (see Section 3.2 of Ref. [37] for details). As a comparison, the SOC strength in Cr$_2$O$_3$ [60], another prototypical AFM insulator, is only around 4 meV, as both Cr and O are light elements. This clearly demonstrates the advantage of using MnBi$_2$Te$_4$ for transduction. In the following, we will take $\lambda_{SOC} = 10 \text{ meV}$ as a conservative estimate.

The strong SOC in MnBi$_2$Te$_4$ also leads to band inversion and nontrivial topology [27–29]. The electronic band structure of MnBi$_2$Te$_4$ is shown in Figure 2b. The inverted bandgap at the Γ point is around 0.2 eV, which falls in the mid-infrared range. The optical responses of MnBi$_2$Te$_4$ have been studied before [17,61–63]. Here we focus on the optical susceptibility $\chi_r(\omega_{IR})$, which can be expressed as

$$\chi_r(\omega_{IR}) = \frac{e^2}{\varepsilon_0} \int_k \frac{d\mathbf{k}}{(2\pi)^3} \sum_{c,v} \frac{|\langle 0,v|r|0,c\rangle|^2}{E_{|0,c\rangle} - E_{|0,v\rangle} - \omega_{IR} + i\zeta}, \quad (3)$$

where $\int d\mathbf{k}$ indicates the integration in the first Brillouin zone, while $c$ ($v$) indicates conduction (valence) bands. $\zeta$ is the linewidth of electronic orbital states, which is uniformly taken as 10 meV. The influence of $\zeta$ is negligible when $\omega_{IR}$ is far from the bandgap of MnBi$_2$Te$_4$. One can see that the magnitude of the numerator, $|\langle 0,v|r|0,c\rangle|^2$, strongly depends on the wavefunction overlap between valence and conduction bands, which is enhanced in topological materials because of the band inversion [23]. The calculated $\chi_r(\omega_{IR})$ of MnBi$_2$Te$_4$ is shown in Figure 3c, where one can see that $\chi_r(\omega_{IR} \to 0)$ is greater than 60, significantly larger than those of typical topologically trivial insulators. As an example, $\chi_r$ is below 15 in Cr$_2$O$_3$ for below-bandgap frequencies (Figure S3 in Ref. [37]). This corroborates the enhanced bulk optical responses in topological insulators [23–26].

To compare the coupling strength $\mathcal{G}$ in Eq. (1) with the conventional definition of second-order susceptibility, we convert $\mathcal{B}$ to $\mathcal{E}$ using $\mathcal{B} = \mathcal{E}/c_0$ with $c_0$ the speed of light, yielding $\mathcal{G} = \chi_{eff}^{(2)} V_0 \varepsilon_0 \mathcal{E}^3$. The effective second-order susceptibility is found to be $\chi_{eff}^{(2)} \gtrsim 10^6$ pm/V in MnBi$_2$Te$_4$ when $\Delta_{MW} = \Delta_{SOC} = 100$ GHz. As a comparison, in electro-optical materials, the second-order susceptibility is typically below $10^3$ pm/V [13]. The large intrinsic nonlinearities also indicate that MTIs can be efficient in other applications besides quantum transduction, such as two-mode squeezing [64] and entanglement generation [65,66]. As explained below, quantum transduction uses the beam-splitting terms in the Hamiltonian. Conversely, two-mode squeezing and entanglement generation typically exploit the squeezing terms. The beam-splitting and two-mode squeezing terms should be equally large in MnBi$_2$Te$_4$, as they have the same microscopic origin described by Eq. (1).

**Experimental realization of the MnBi$_2$Te$_4$ transducer.** In this section, we discuss several issues relevant to the experimental realization of quantum transduction based on MnBi$_2$Te$_4$. The goal is to achieve efficient conversion between MW and IR photons. Hence, we consider a setup where the MnBi$_2$Te$_4$ sample is coupled to both an IR cavity and a MW cavity. Additionally, a classical IR pumping laser with a frequency $\omega_{pump}$ is employed to satisfy the frequency-matching condition ($\omega_{MW} + \omega_{pump} = \omega_{IR}$). Accordingly, in Eq. (1), we second quantize the magnetic field $\mathcal{B} \equiv \mathcal{B}_{MW}^{zpf}(a_{MW}^\dagger + a_{MW})$ and electric field $\mathcal{E}^q \equiv \mathcal{E}_{IR}^{zpf}(a_{IR}^\dagger + a_{IR})$, and treat $\mathcal{E}^p \equiv \mathcal{E}_{pump}$ as the classical pumping field. Here $a_{MW\,(IR)}^\dagger$ is the creation operator for the MW (IR) photon, and $a_{\text{MW (IR)}}$ is the corresponding annihilation operator. $\mathcal{B}_{\text{MW}}^{\text{zpf}} \equiv \sqrt{\frac{\mu_0 \omega_{\text{MW}}}{2V_{\text{MW}}}}$ and $\mathcal{E}_{\text{IR}}^{\text{zpf}} \equiv \sqrt{\frac{\omega_{\text{IR}}}{2\varepsilon_0(\chi_r+1)V_{\text{IR}}}}$ are the zero-point magnetic and electric fields of the MW and IR cavities, whose mode volumes are $V_{\text{MW}}$ and $V_{\text{IR}}$, respectively. $\mu_0$ is the vacuum permeability.

For the transduction process, we need to focus on the beam-splitting terms in Eq. (1), which can be expressed as $\mathcal{H} = \mathcal{G}(a_{\text{MW}}^\dagger a_{\text{IR}} + a_{\text{IR}}^\dagger a_{\text{MW}})$. The overall coupling strength is

$$\mathcal{G} = F \cdot \frac{\rho g_e \mu_B S \cdot \lambda_{\text{SOC}} \cdot \varepsilon_0 \chi_r V_{\text{cell}}}{\Delta_m \Delta_{\text{SOC}}} \cdot \sqrt{\frac{\mu_0 \omega_{\text{MW}} \omega_{\text{IR}}}{4\varepsilon_0(\chi_r+1)}} \cdot \mathcal{E}_{\text{pump}} \quad (4)$$

where $\rho$ is the density of magnetic moments. Compared with Eq. (1), here we introduce an additional filling factor

$$F \equiv \frac{1}{\sqrt{V_{\text{IR}} V_{\text{MW}}}} \int_{V_0} \phi_{\text{pump}}(r) \phi_{\text{IR}}(r) \phi_{\text{MW}}(r) dV \quad (5)$$

where $V_0$ is the volume of the MnBi$_2$Te$_4$ crystal, while $\phi_{\text{pump}}(r)$, $\phi_{\text{IR}}(r)$ and $\phi_{\text{MW}}(r)$ are the mode functions of the pump laser, the IR, and the MW cavities, respectively. Notably, $F$ characterizes the imperfect mode overlap of the three fields within the crystal volume. To improve the overall coupling strength $\mathcal{G}$, large $F$ is desirable.

Here we briefly discuss possible designs of the IR and MW cavities. Using a simple confocal IR cavity, a mode volume below $1 \text{ mm}^3$ can be readily realized. For example, one has $V_{\text{IR}} = 0.1 \text{ mm}^3$ when the length and mirror radius of the cavity are $L = R \approx 3 \text{ cm}$. Experimentally, much smaller mode volumes have been realized [67]. The intrinsic cavity loss comes mainly from defect absorption, which is around $\kappa_{\text{IR}}^i \approx 0.3 \text{ GHz}$ when the defect concentration is 1ppm (Section 2 of Ref. [37]). As we will discuss below, the overall transduction efficiency is proportional to the ratio $\frac{\kappa_{\text{IR}}^c}{\kappa_{\text{IR}}^i + \kappa_{\text{IR}}^c}$, where $\kappa_{\text{IR}}^c$ is the coupling loss of the IR cavity. In this regard, we assume one of the two cavity mirrors has a very low reflectivity $r_1$. Specifically, one has $\kappa_{\text{IR}}^c \approx 4.5 \text{ GHz}$ with $r_1 = 0.1$ and $L = 3 \text{ cm}$. Under these assumptions, the quality factor of the cavity is on the order of $10^5$, which is a moderate value.

Several different designs of MW cavities have been proposed to facilitate the integration with the IR cavity. This ranges from simple loop-gap resonators [3,10] to cavities with more sophisticated geometries [68]. The MW cavity loss mostly comes from the external coupling $\kappa_{MW}^c$, instead of intrinsic loss $\kappa_{MW}^i$, so we will take $\frac{\kappa_{MW}^c}{\kappa_{MW}^c+\kappa_{MW}^i} \approx 1$ in the following. These designs result in $F$ ranging from 0.0084 in Ref. [3] to above 0.1 in Ref. [68]. Potentially, $F$ can be further improved by optimizing the geometries of the MW and IR cavities, and we will leave this for future work. Here we take a conservative value of $F = 0.01$.

Taking $\omega_{MW} = 10$ GHz, $\omega_{IR} = 0.1$ eV, and $\Delta_m = \Delta_{SOC} = 100$ GHz, one has $\mathcal{G}[\text{GHz}] \sim 1.0 \times \mathcal{E}_{pump}[\text{MV} \cdot \text{m}^{-1}]$. Hence, one has $\mathcal{G} = 0.1$ GHz with $\mathcal{E}_{pump} = 0.1$ MV$\cdot$m$^{-1}$, corresponding to a laser intensity of $13$ W$\cdot$mm$^{-2}$. This is a relatively weak pump. Particularly, the temperature rise due to heat absorption from the pump can be kept far below 1 mK when a pulsed pump is used (Section 2 of Ref. [37]). Such a temperature rise is marginal, since we expect the experiments to be performed at a temperature of $10 \sim 100$ mK. This is because the microwave frequency is around 10 GHz, equivalent to 500 mK. Hence, the thermal occupation of the microwave mode can be kept below $10^{-2}$ when the temperature is $10 \sim 100$ mK. Additionally, the Néel temperature of MnBi$_2$Te$_4$ is tens of Kelvin, hence a temperature of 100 mK is sufficient to sustain the magnetic order of MnBi$_2$Te$_4$.

**Performance of quantum transduction based on MnBi$_2$Te$_4$.** Using the input-output formalism, the quantum transduction efficiency $\eta$ can be expressed as [3,68,69]

$$\eta(\Delta_{IR}) = \frac{\kappa_{IR}^c}{\kappa_{IR}^0} \frac{\kappa_{MW}^c}{\kappa_{MW}^0} \left| \frac{4\mathcal{G}\sqrt{\kappa_{MW}^0 \kappa_{IR}^0}}{4\mathcal{G}^2 + (\kappa_{MW}^0 - 2i\Delta_{IR})(\kappa_{IR}^0 - 2i\Delta_{IR})} \right|^2 \tag{6}$$

where $\Delta_{IR} \equiv \omega_{MW} + \omega_{pump} - \omega_{IR}$ is the detuning from frequency matching condition, while $\kappa_{MW\,(IR)}^0 = \kappa_{MW\,(IR)}^i + \kappa_{MW\,(IR)}^c$ is the total linewidth. For the IR cavity, we first take $\kappa_{IR}^i = 0.3$ GHz and $\kappa_{IR}^c = 4.5$ GHz, as described before. In this case, the impedance matching condition $\Gamma = \frac{2\mathcal{G}}{\sqrt{\kappa_{IR}^0 \kappa_{MW}^0}} = 1$ can be realized with $\kappa_{MW}^0 = 8.3$ MHz, corresponding to a quality factor of 1200 for the MW cavity, which is a moderate requirement [70,71]. Using these values, the overall conversion efficiency is $\eta = \frac{\kappa_{IR}^c}{\kappa_{IR}^0} \approx 94\%$ at $\Delta_{IR} = 0$.

We can rewrite the transduction efficiency $\eta$ as a function of the two cavity quality factors, $Q_{MW}^0$ and $Q_{IR}^0$. For the MW cavity, we assume the intrinsic loss is negligible, so $\kappa_{MW}^0 \approx \kappa_{MW}^c$ and the quality factor is $Q_{MW}^0 = \frac{\omega_{MW}}{\kappa_{MW}^c}$. For the IR cavity, we fix $\kappa_{IR}^i = 0.3$ GHz, yielding $Q_{IR}^0 = \frac{\omega_{IR}}{\kappa_{IR}^c + 0.3 \text{ GHz}}$. We also fix $\mathcal{G} \approx 0.1$ GHz. The maximum transduction efficiency and full width at half maximum (FWHM) bandwidth are shown in Figure 3. One can see that $\eta > 0.9$ can be realized with a relatively wide range of parameters. Moreover, the maximum FWHM bandwidth is $2\sqrt{2}\mathcal{G}$, according to the input-output formalism. This again highlights the importance of using materials with large nonlinearities.

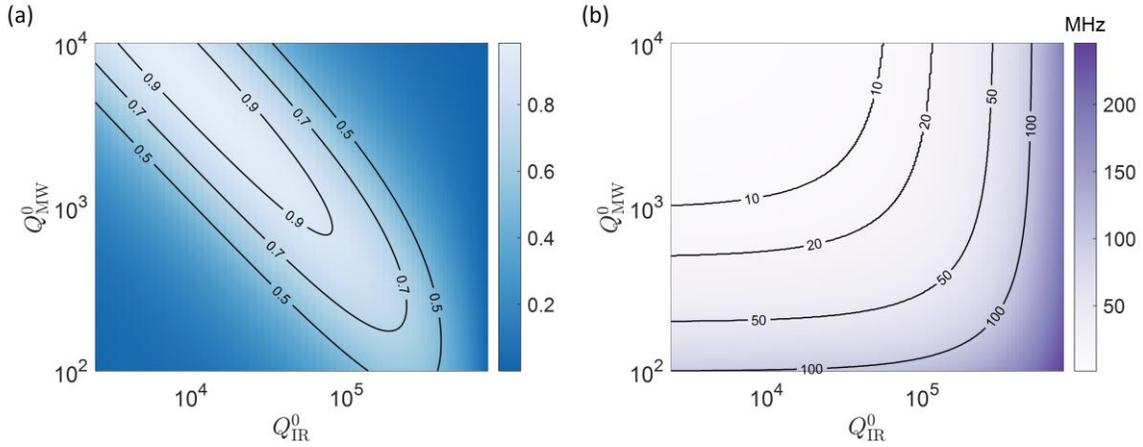

**Figure 3**. (a) Maximum transduction efficiency $\eta_{max}$ at $\Delta_{IR} = 0$ and (b) bandwidth ($\Delta_{IR}$ at $\frac{\eta = \eta_{max}}{2}$) predicted from the input-output formalism. The solid lines are contours of constant values.

**Discussions**. Quantum transduction is a challenging task, and several crucial factors can hinder efficient quantum transduction, including the necessity for high-quality IR/MW cavities and the heating problem caused by optical pumps. Specifically, the transduction efficiency is controlled by the phase-matching parameter $\Gamma = \frac{2\mathcal{G}}{\sqrt{\kappa_{IR}^0 \kappa_{MW}^0}}$, and the phase matching condition $\Gamma = 1$ requires $\mathcal{G}$ to be comparable with $\kappa_{IR}^0$ and $\kappa_{MW}^0$. This is challenging as the nonlinearities are usually rather weak. Hence, typically one needs to use (1) strong optical pump, which can lead to the heating problem and possible sample damage; and (2) optical/microwave cavities with small linewidth and large quality factors. This also limits the transduction bandwidth, which is comparable with the cavity linewidth.

In this regard, materials with large nonlinearity have significant implications for improving the efficiency of quantum transduction. It is actually the key to resolving many of the challenges described above. Thanks to the large intrinsic nonlinearity of MnBi$_2$Te$_4$, $\mathcal{G} \gtrsim 1$ GHz can be realized with relatively weak optical pumps, which can significantly mitigate the heating problem. The GHz coupling strength also indicates that the transduction process can be completed within nanoseconds. Hence, a nanosecond pulsed optical pump is sufficient, which can further suppress the total heat absorption (Section 2 of Ref. [37]). Furthermore, a GHz coupling strength also implies that the linewidth $\kappa_{\text{IR}}^0$ and $\kappa_{\text{MW}}^0$ can be on the order of GHz as well, which is much less demanding than the requirements of $\kappa_{\text{IR}}^0, \kappa_{\text{MW}}^0 \sim$ MHz in many previous proposals [3,68]. This can not only make the device fabrication much easier, but also enlarge the transduction bandwidth significantly.

As a comparison, the nonlinearities of topologically trivial magnetic materials, such as Cr$_2$O$_3$, or conventional electro-optical materials, are usually two to three orders of magnitude lower than that of MnBi$_2$Te$_4$. This indicates that if the same pumping laser is used, then $\mathcal{G}$ would be much smaller, leading to extremely low transduction efficiency (below $10^{-4}$). Conversely, to achieve the same transduction performance, then $\mathcal{E}_{\text{pump}}$ must be $10^2 \sim 10^3$ times stronger in Cr$_2$O$_3$ or conventional electro-optical materials. This could cause significant heating, photon relaxation, and damage in the materials, which can be catastrophic for the transduction process. As an example, the heating from defect states and two-photon absorption scale as $\mathcal{E}_{\text{pump}}^2$ and $\mathcal{E}_{\text{pump}}^4$, respectively, so the temperature rise can reach hundreds of Kelvins when $\mathcal{E}_{\text{pump}}$ is increased to $10$ MV $\cdot$ m$^{-1}$ (Section 2 of Ref. [37]). This again highlights the advantages of using MnBi$_2$Te$_4$ as the transducer, which enjoys exceptionally high intrinsic nonlinearities.

Our work also unveils an important property of topological materials, namely the pronounced bulk optical nonlinearities. Until now, we considered quantum transduction to exemplify the potential of MTIs in quantum engineering. More broadly, MTIs can be advantageous in many other applications that require strong linear or nonlinear optical responses, such as quantum squeezing [64], entanglement generation [65,66], spontaneous parametric down-conversion [72], etc. It is worth mentioning that the energy gap of $MnBi_2Te_4$, which is necessary for quantum transduction and has been predicted in theories, remains elusive until very recent experiments [73]. Fortunately, the large nonlinearities are not exclusive to $MnBi_2Te_4$; they should extend to other magnetic topological insulators as well, and a catalog of magnetic topological insulators is available [14]. This can offer more opportunities to achieve efficient quantum transduction by using magnetic topological insulators. Moreover, it is also possible to use non-topological magnetic materials. For example, $MnPSe_3$ [74] can be a promising candidate thanks to its low magnon frequency, leading to smaller detuning between magnons and infrared photons.

Furthermore, $MnBi_2Te_4$ can be thinned down to single or multiple layers [27], leading to the possibility of quantum applications in the two-dimensional limit [75]. Experimentally, both bulk samples of $MnBi_2Te_4$ with millimeter size [29] and two-dimensional samples with a tunable number of layers [16,76] have been fabricated. Sophisticated devices involving $MnBi_2Te_4$ have also been demonstrated experimentally [77,78] as well. To the best of our knowledge, there have been no prior works that integrate $MnBi_2Te_4$ with quantum transduction devices. We hope our work can draw the attention of materials and device scientists toward exploring this promising direction.

In conclusion, we proposed that MTIs, such as $MnBi_2Te_4$, can serve as efficient quantum transducers, thanks to their strong intrinsic optical nonlinearities. This is primarily due to topologically enhanced optical responses, robust spin-orbit coupling, together with high spin density in MTIs. We demonstrate that transduction efficiency over 90% and transduction bandwidth over GHz can be simultaneously achieved with modest experimental requirements, which can facilitate diverse applications in quantum information science and other fields.

## Acknowledgment

We acknowledge support from the Office of Naval Research MURI through grant #N00014-17-1-2661 and Honda Research Institute USA.